\documentclass[prb,aps,superscriptaddress,floatfix,tightenlines,showpacs,twocolumn]{revtex4-2}
\usepackage{graphicx}
\usepackage{dcolumn}   
\usepackage{bm}        
\usepackage{amssymb}   
\usepackage{amsmath}
\usepackage{url}
\usepackage{layout}
\usepackage{color}
\usepackage{epsfig}
\usepackage{graphicx}
\usepackage{mathrsfs}\usepackage{bm}\usepackage{appendix}\usepackage{color}
\usepackage{makecell}
\usepackage{booktabs}
\usepackage{float}
\usepackage[hyperindex,breaklinks]{hyperref}

\def\be{\begin{equation}}       \def\ee{\end{equation}}
\def\bea{\begin{eqnarray}}      \def\eea{\end{eqnarray}}
\def\bp{\begin{pmatrix}} \def\ep{\end{pmatrix}}
\def\beaa{\begin{equation}\begin{aligned}}
		\def\eeaa{\end{aligned}\end{equation}}

\begin{document}

\title{Effect of Rare-earth Element Substitution in Superconducting R$_3$Ni$_2$O$_7$ Under Pressure}

\author{Zhiming Pan}
\thanks{These two authors contributed equally to this work.}
\affiliation{Institute for Theoretical Sciences, Westlake University, Hangzhou 310024, Zhejiang, China}
\affiliation{New Cornerstone Science Laboratory, Department of Physics, School of Science, Westlake University, Hangzhou 310024, Zhejiang, China}
\author{Chen Lu}
\thanks{These two authors contributed equally to this work.}
\affiliation{New Cornerstone Science Laboratory, Department of Physics, School of Science, Westlake University, Hangzhou 310024, Zhejiang, China}
\author{Fan Yang}
\email{yangfan\_blg@bit.edu.cn}
\affiliation{School of Physics, Beijing Institute of Technology, Beijing 100081, China}
\author{Congjun Wu}
\email{wucongjun@westlake.edu.cn}
\affiliation{New Cornerstone Science Laboratory, Department of Physics, School of Science, Westlake University, Hangzhou 310024, Zhejiang, China}
\affiliation{Institute for Theoretical Sciences, Westlake University, Hangzhou 310024, Zhejiang, China}
\affiliation{Key Laboratory for Quantum Materials of Zhejiang Province, School of Science, Westlake University, Hangzhou 310024, Zhejiang, China}
\affiliation{Institute of Natural Sciences, Westlake Institute for Advanced Study, Hangzhou 310024, Zhejiang, China}

\begin{abstract}
Recently, high temperature ($T_c\approx 80$K) superconductivity (SC) has been discovered in La$_3$Ni$_2$O$_7$ (LNO) under pressure.
Question arises whether the transition temperature $T_c$ could be further enhanced under suitable conditions.
A possible route for realizing higher $T_c$ is element substitution. 
Similar SC could appear in rare-earth (RE) R$_3$Ni$_2$O$_7$ (RNO, R=RE element) material series under pressure.
The electronic properties in the RNO materials are dominated by the Ni $3d$ orbitals in the bilayer NiO$_2$ plane.
In the strong coupling limit, the SC could be fully characterized by a bilayer single $3d_{x^2-y^2}$-orbital $t$-$J_{\parallel}$-$J_{\perp}$ model.
Under RE element substitution from La to RE element, the lattice constant decreases and the electronic hopping increases, leading to stronger superexchanges between the $3d_{x^2-y^2}$ orbitals.
Based on the slave-boson mean-field theory, we explore the pairing nature and the evolution of $T_c$ in RNO materials. Consequently, it is found that the element substitution does not alter the pairing nature, i.e. the inter-layer $s$-wave pairing is always favored in RNO. 
However, the $T_c$ increases from La to Sm and a nearly doubled $T_c$ is achieved for SmNO.
This work provides evidence for possible higher $T_c$ R$_3$Ni$_2$O$_7$ materials, which may be realized in further experiments.
\end{abstract}
\maketitle


{\bf Introduction} The recent discovered high-temperature superconductivity in La$_3$Ni$_2$O$_7$ (LNO) ~\cite{Wang2023LNO} has attracted great interests both experimentally ~\cite{WenHH2023,Wang2023LNOb,YuanHQ2023LNO,yang2023arpes,zhang2023pressure} and theoretically ~\cite{YaoDX2023,Dagotto2023,WangQH2023,lechermann2023,Kuroki2023,HuJP2023,ZhangGM2023DMRG,Werner2023,shilenko2023correlated,WuWei2023charge,cao2023flat,chen2023critical,YangF2023,lu2023bilayertJ,oh2023type2,zhang2023structural,liao2023electron,qu2023bilayer,Yi_Feng2023,jiang2023high,zhang2023trends,huang2023impurity,qin2023high,tian2023correlation,jiang2023pressure,lu2023sc,luo2023high}.
The signature of the superconducting transition temperature $T_c$ is about $80$K under pressures over $14$GPa ~\cite{Wang2023LNO}, 
which manifests as a newly platform of studying high-$T_c$ superconductor other than cuprates~\cite{bednorz1986LBCO,proust2019highTc}.

The basic ingredient for the electronic properties in LNO is the bilayer NiO$_2$ planes \cite{Wang2023LNO}.
On average, each Ni$^{2.5+}$ is half-filled in its $3d_{z^2}$ orbital and quarter-filled in its $3d_{x^2-y^2}$ orbital.
The spin alignment in the two $E_g$ orbitals is further restricted from the Hund's rule.
The $3d_{z^2}$ electrons could hop between the two NiO$_2$ layers through the intermediate O$2p_z$ orbitals,
while the $3d_{x^2-y^2}$ electrons mostly hop within the layers.
The physical filling of the two $E_g$ orbitals could deviate due to the self-doped effect from hybridization between them \cite{Yi_Feng2023,zhang2023trends} or doping of holes on the O$2p$ orbitals\cite{WuWei2023charge}.

Taken into account the Hund's rule, we have suggested an effective bilayer single-$3d_{x^2-y^2}$ orbital $t$-$J_{\parallel}$-$J_{\perp}$ model as the minimal model for the high-$T_c$ superconductivity in the strong coupling limit \cite{lu2023bilayertJ}.
Each NiO$_2$ layer is composed by a conventional $t$-$J_{\parallel}$ model with nearest-neighbor intra-layer antiferromagnetic (AFM) spin exchange $J_{\parallel}$.
The two NiO$_2$ layers couple through an effective inter-layer AFM spin exchange $J_{\perp}$ between $3d_{z^2}$ electrons lying on the two layers,
which is generated from integrating out the $3d_{z^2}$ orbital degrees of freedom under the strong Hund's coupling.
The superconducting transition temperature could be dramatically enhanced when the inter-layer coupling $J_{\perp}$ is larger than the intra-layer one\cite{lu2023bilayertJ}. 
This kind of bilayer $t$-$J$ model has already been explored in multilayer cuprates \cite{ubbens1994,kuboki1995,maly1996,nazarenko1996bilayer3d,medhi2009} and may be realized recently in ultracold atoms \cite{bohrdt2021exploration,demler2022,Grusdt2023}.
Most of these works focus on the theoretical side with different physical parameter regime \cite{ubbens1994,kuboki1995,maly1996,nazarenko1996bilayer3d,medhi2009,demler2022,eder1995,vojta1999,zhao2005,zegrodnik2017,bohrdt2021exploration,demler2022,Grusdt2023}.
The strong interlayer coupling $J_{\perp}$ in LNO plays important role for high $T_c$\cite{lu2023bilayertJ}, 
which is hard to be realized in multilayer cuprates.

Effect of element substitution is an important chemical approach to increase the superconducting transition temperature.
Replacing the La element by other rare-earth (RE) elements could influence the crystal and electronic structure of the materials.
Elements substitution alters the lattice constant, modifying the hopping strength of electrons.
Under suitable pressure, R$_3$Ni$_2$O$_7$ (RNO, R=RE elements) material series undergo similar structure transition to the Fmmm phase as LNO \cite{zhang2023trends}, opening up the possibility of realizing high $T_c$ in these materials.

Previous first principle density functional theory (DFT) for R$_3$Ni$_2$O$_7$ reveals the RE elements dependence of the physical parameters\cite{zhang2023trends}.
The hopping strength could be enhanced through element substitution.
However, the weak coupling analysis based on random phase approximation (RPA) predicts that the pairing strength is reduced from La to Sm, 
and the superconducting transition temperature $T_c$ decreases simultaneously \cite{zhang2023trends}.
Such calculation might suggest that LNO already has the largest $T_c$ in the RNO material series.

The situation could be reversed in the strong coupling picture, where $T_c$ is related to the superexchange strength $J_{\perp}$\cite{lu2023bilayertJ}.
Based on the hopping strength from the DFT calculations \cite{zhang2023trends}, the effective AFM couplings $J_{\perp}$ and $J_{\parallel}$ increase simultaneously from La to Sm. 
Previous calculation has already seen that strong super-exchange $J_{\perp}$ could enhance the $T_c$, leading to the high-$T_c$ observed in the LNO experiment \cite{lu2023bilayertJ}.
All the relevant energy scales grow up under the element substitution.
These observations provide potential increased $T_c$ under elements substitution.

In this work, we study the effect of RE element substitution in RNO materials based on the effective bilayer single $3d_{x^2-y^2}$-orbital $t$-$J_{\parallel}$-$J_{\perp}$ model\cite{lu2023bilayertJ}.
We first obtain the effective AFM spin exchange $J_{\perp}$ and $J_{\parallel}$ for different RE elements from the DFT calculation\cite{zhang2023trends}.
Then, we apply the slave-boson mean field analysis \cite{kotliar1988,lee2006htsc} for the bilayer $t$-$J$ model within these physical parameters.
The superconducting pairing gaps and transition temperatures $T_c$ are numerical calculated. Our results suggest that the element substitution does not change the pairing nature, i.e. the inter-layer s-wave pairing. However, the $T_c$ could be strongly enhanced through element substitution in this strong coupling scenario,
and especially from La to Sm, $T_c$ nearly doubles.
This work provides possible route for enhancing $T_c$ and appeals for experimental verification.

{\bf Effective bilayer model}
In the double-layered Ruddlesden-Popper R$_3$Ni$_2$O$_7$ material, 
electronic properties are decoded in the bilayer NiO$_2$ planes.
DFT calculation \cite{pardo2011dft,Wang2023LNO,zhang2023trends} indicates that the relevant physical degrees of freedom come from the two $E_g$ orbital in Ni$^{2.5+}$, 
where $3d_{z^2}$ orbital is half-filled and $3d_{x^2-y^2}$ orbital is quarter-filled.
The inter-layer hopping $t_{zz}^{\perp}$ between the $3d_{z^2}$ orbitals and the intra-layer hopping $t_{xx}^{\parallel}$ between the $3d_{x^2-y^2}$ orbitals dominate the mobile electron process.
In the strong coupling limit with large Hubbard-$U$ ($U\simeq 4$eV), an effective inter-layer spin superexchange for $3d_{z^2}$ orbitals and intra-layer one for $3d_{x^2-y^2}$ orbitals are generated,
\begin{equation}
J_{\perp}=\frac{4t_{zz}^{\perp 2}}{U},\quad
J_{\parallel}=\frac{4t_{xx}^{\parallel 2}}{U}.
\label{eq:Jcoupling}
\end{equation}
Moreover, $J_{\perp}$ is larger than $J_{\parallel}$ due to the larger inter-layer hopping of $3d_{z^2}$ orbitals.

\begin{figure}[t]
\includegraphics[width=0.7\linewidth]{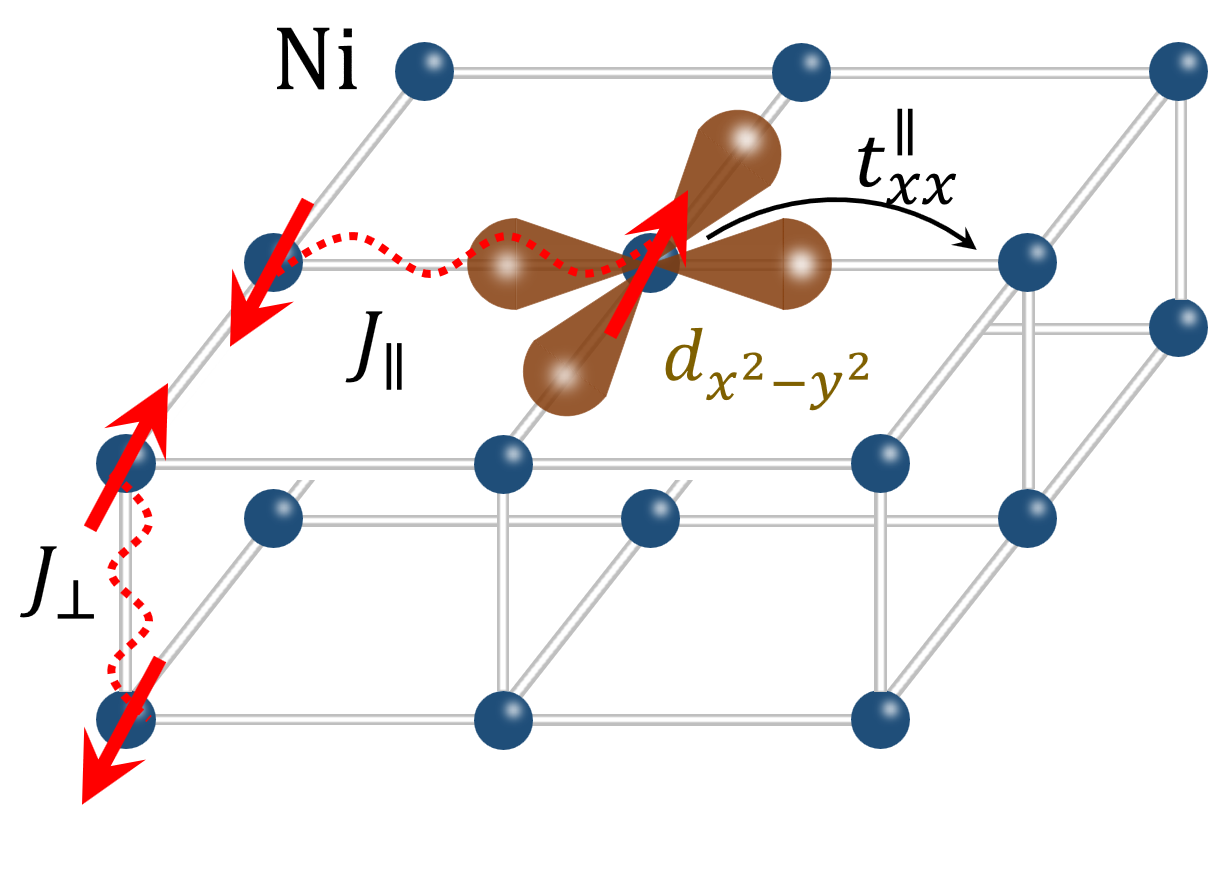}
\caption{Schematic diagram for the effective single $3d_{x^2-y^2}$ orbital bilayer $t$-$J_{\parallel}$-$J_{\perp}$ model.
Each layer is comprised by a conventional intra-layer $t$-$J_{\parallel}$ model, while the two layers interact through an inter-layer spin exchange $J_{\perp}$.  
}
\label{fig:Nit-JModel}
\end{figure}

The relevant physics in RNO is described by a bilayer two-orbital $t$-$J$-$J_H$ model,
\begin{align}
H=&-t_{xx}^{\parallel} \sum_{\langle i,j\rangle \alpha \sigma} \big(d_{x^2i\alpha\sigma}^{\dagger} d_{x^2j\alpha\sigma} +\text{h.c.}\big)  \notag\\
&+J_{\parallel} \sum_{\langle i,j\rangle \alpha} \bm{S}_{x^2i\alpha} \cdot \bm{S}_{x^2j\alpha} \notag\\
&+J_{\perp} \sum_{i} \bm{S}_{z^2i1} \cdot\bm{S}_{z^2i2}
-J_H\sum_{i\alpha} \bm{S}_{x^2i\alpha}\cdot \bm{S}_{z^2i\alpha},
\label{eq:tJJHmodel}
\end{align}
where $d_{x^2i\alpha\sigma}^{\dagger}$ is electron creation operator for $3d_{x^2-y^2}$ orbital at the lattice site $i$, $\alpha=1,2$ is the layer index for the two NiO plane and $\sigma=\uparrow,\downarrow$ is the spin index.
$\bm{S}_{x^2i\alpha}=\frac{1}{2}d_{x^2i\alpha}^{\dagger}[\bm{\sigma}]d_{x^2i\alpha}$ is the spin operator for $3d_{x^2-y^2}$ electron, with Pauli matrix $\bm{\sigma}=(\sigma_x,\sigma_y,\sigma_z)$.
The summation $\sum_{\langle i,j\rangle}$ takes over all the nearest neighbor sites $i,j$ within the plane.
$S_{z^2i\alpha}$ is the spin operator for the localized $3d_{z^2}$ orbital.
The two orbitals interact through the strong on-site Hund's coupling $J_H$.

The effective minimal model under strong Hund's coupling is a bilayer $t$-$J_{\parallel}$-$J_{\perp}$ model for the single $3d_{x^2-y^2}$ orbital\cite{lu2023bilayertJ}. 
The strong Hund's coupling binds the spins of $3d_{x^2-y^2}$ and $3d_{z^2}$ orbitals into a spin-triplet state.
The inter-layer $3d_{z^2}$ spin exchange $J_{\perp}$ will be transmitted to $3d_{x^2-y^2}$ spins,
generating an effective inter-layer spin exchange $J_{\perp}$ between $3d_{x^2-y^2}$ spins.
Integrating out the $3d_{z^2}$ degrees of freedom under strong Hund's coupling ($J_H\gg J_{\perp}$), the two-orbital system Eq.~ (\ref{eq:tJJHmodel}) reduces to the bilayer single $3d_{x^2-y^2}$-orbital model, 
\begin{equation}
\begin{aligned}
H=&-t_{xx}^{\parallel} \sum_{\langle i,j\rangle \alpha\sigma} \big(d_{x^2i\alpha\sigma}^{\dagger} d_{x^2j\alpha\sigma} +\text{h.c.}\big) 
 \\
&-t_{xx}^{\perp} \sum_{i\sigma} \big(d_{x^2i1\sigma}^{\dagger} d_{x^2i2\sigma} +\text{h.c.}\big) 
 \\
&+J_{\parallel} \sum_{\langle i,j\rangle \alpha} \bm{S}_{x^2i\alpha} \cdot \bm{S}_{x^2j\alpha} +J_{\perp} \sum_{i} \bm{S}_{x^2i1} \cdot \bm{S}_{x^2i2},
\end{aligned}
\label{eq:x2-tJ-H}
\end{equation}
with intra-layer nearest-neighbor AFM spin exchanges $J_{\parallel}$ and inter-layer one $J_{\perp}$, as depicted in Fig.~(\ref{fig:Nit-JModel}).
A small inter-layer hopping $t_{xx}^{\perp}$ is added, which could fix the relative phases between the pairings of the two layers.
In the reduced model, $3d_{z^2}$ orbital plays as an intermediate hidden bridge for the inter-layer coupling $J_{\perp}$ between $3d_{x^2-y^2}$ orbitals.
It also accounts for the self-doping effect in the $3d_{x^2-y^2}$ orbital \cite{Yi_Feng2023,zhang2023trends}.

The filling level $x$ (or doping level $\delta=1-2x$) of $3d_{x^2-y^2}$ orbital deviates from quarter filling due to the hybridization between the two $E_g$ orbitals.
$C_4$ rotation symmetry around the $z$-axis forbids the on-site hybridization between $3d_{z^2}$ and $3d_{x^2-y^2}$ orbital.
However, symmetry arguments still persist a finite hybridization between neighbor sites.
In the DFT calculation\cite{YaoDX2023,zhang2023trends}, such nearest-neighbor hybridization between $3d_{z^2}$ and $3d_{x^2-y^2}$ orbital has been shown to be comparable to the nearest-neighbor hopping in the RNO materials.
The two orbitals mix together, leading to bonding or anti-bonding state. 
Moreover, an effective intra-layer hopping process could be generated for $3d_{z^2}$ orbital.
The averaged densities of the $3d_{z^2}$ and $3d_{x^2-y^2}$ orbitals deviate from half filling and quarter filling owing to the density fluctuations.
For the RE elements from La to Sm, $3d_{x^2-y^2}$ orbital nearly keeps the same physical filling $x\approx 0.3$ from the DFT calculation \cite{zhang2023trends}.

\begin{figure}[t]
\includegraphics[width=0.7\linewidth]{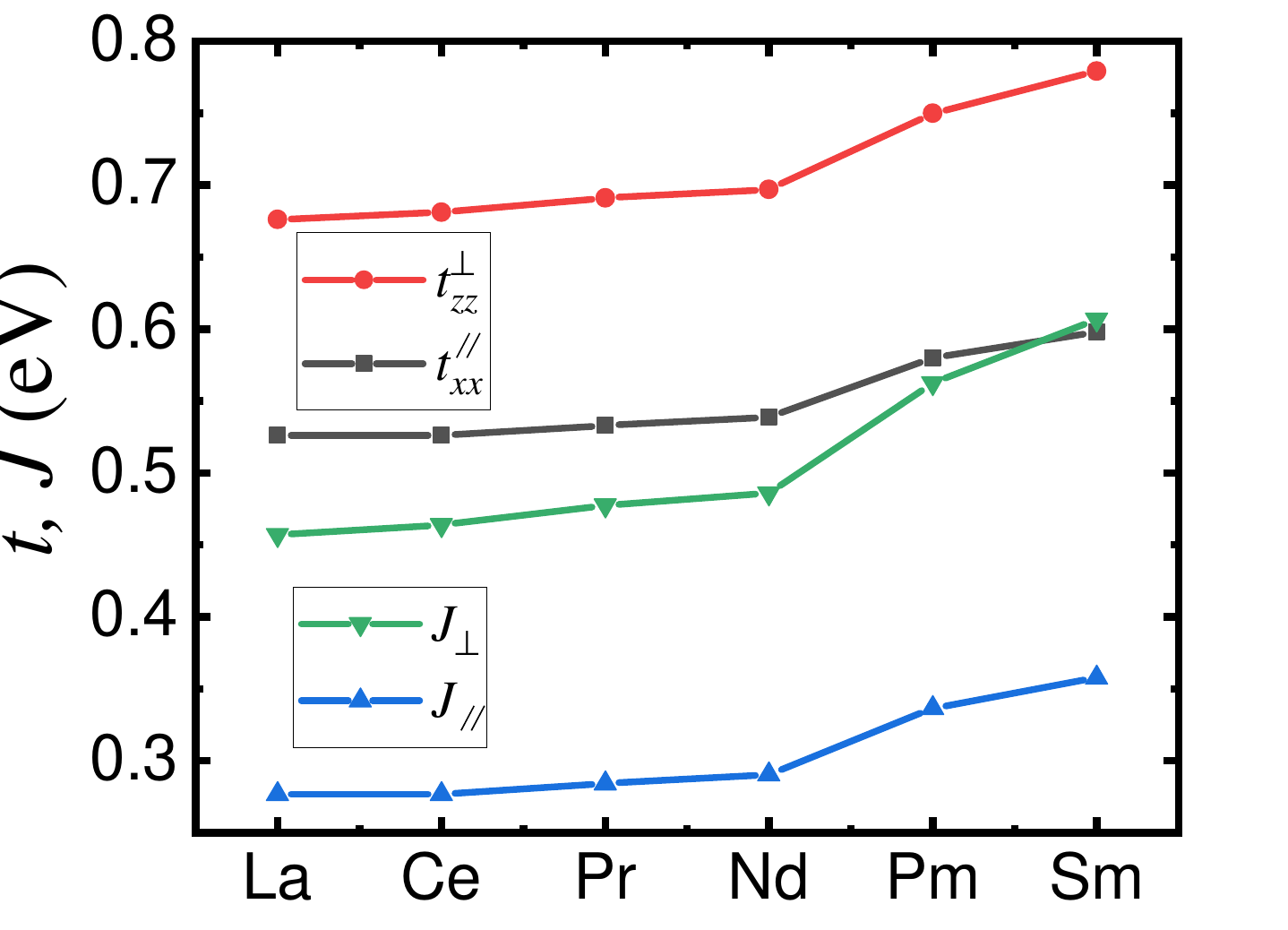}
\caption{Relevant hopping strength and effective spin exchange for different RE elements from La to Sm.
From La to Sm, the inter-layer $3d_{z^2}$ hopping $t_{zz}^{\perp}$ and intra-layer $3d_{x^2-y^2}$ hopping $t_{xx}^{\parallel}$ both gradually increase \cite{zhang2023trends}.
The effective inter-layer $3d_{z^2}$ spin exchange $J_{\perp}$ and intra-layer $3d_{x^2-y^2}$ spin exchange $J_{\parallel}$ follow the similar tendency. 
}
\label{fig:REtJ}
\end{figure}

{\bf Slave boson mean field theory}
In the slave boson mean field theory \cite{kotliar1988,lee2006htsc}, the electron operator is represented as $d_{x^2i\alpha\sigma}^{\dagger}=f_{i\alpha\sigma}^{\dagger} b_{i\alpha}$,
where $f_{i\alpha\sigma}^{\dagger}$ is the spinon creation operator and $b_{i\alpha}$ is the holon annihilation operator.
The super exchange term can be decoupled in the hopping and pairing channel, i.e., for the inter-layer one,
\begin{equation}
\begin{aligned}
&J_{\perp} \bm{S}_{x^2i1} \cdot \bm{S}_{x^2i2}   \\
=&-
\Big[\chi_{i\perp}
\big(f_{i1\uparrow}^{\dagger} f_{i2\uparrow}
+f_{i1\downarrow}^{\dagger} f_{i2\downarrow}\big)
+\text{h.c.}- \frac{8}{3J_{\perp}} |\chi_{i\perp}|^2 \Big]\\ 
&-
\Big[\Delta_{i\perp}
\big(f_{i1\uparrow}^{\dagger}f_{i2\downarrow}^{\dagger} 
-f_{i1\downarrow}^{\dagger}f_{i2\uparrow}^{\dagger} \big) 
+\text{h.c.} -\frac{8}{3J_{\perp}} |\Delta_{i\perp}|^2  \Big],
\end{aligned}
\end{equation}
and similar decomposition for the intra-layer super exchange.
In the mean-field ansatz, the hopping and pairing order parameters are represented by their mean-field values \cite{lu2023bilayertJ},
\begin{equation}
\begin{aligned}
\chi_{ij}^{(\alpha)}
=&\frac{3}{8} J_{\parallel} \langle f_{j\alpha\uparrow}^{\dagger} f_{i\alpha\uparrow}
+f_{j\alpha\downarrow}^{\dagger} f_{i\alpha\downarrow}\rangle
\equiv \chi_{j-i}^{(\alpha)}, \\
\Delta_{ij}^{(\alpha)}
=&\frac{3}{8} J_{\parallel} \langle f_{j\alpha\downarrow} f_{i\alpha\uparrow} 
-f_{j\alpha\uparrow} f_{i\alpha\downarrow} \rangle
\equiv \Delta_{j-i}^{(\alpha)},  \\
\chi_{i\perp}
=&\frac{3}{8} J_{\perp} \langle f_{i2\uparrow}^{\dagger} f_{i1\uparrow}
+f_{i2\downarrow}^{\dagger} f_{i1\downarrow}\rangle
\equiv \chi_{\perp}, \\
\Delta_{i\perp}
=&\frac{3}{8} J_{\perp} \langle f_{i2\downarrow} f_{i1\uparrow} 
-f_{i2\uparrow} f_{i1\downarrow} \rangle
\equiv \Delta_{\perp}.
\end{aligned}
\end{equation}
The holon will condense at low temperature and holon operators can be replaced by the condensation density, $b_{i\alpha}\sim b_{i\alpha}^{\dagger}\sim \sqrt{\delta}=\sqrt{1-2x}$.

The superconducting state is achieved when the spinon is paired and the holon is condensed\cite{kotliar1988}.
There exist two typical temperature scales, $T_{\text{BEC}}$ for the holon condensation, and $T_{\text{BCS}}$ for the spinon pairing.
The doping dependence of them differs from each other.
$T_{\text{BEC}}$ increases almost linearly as the doping level $\delta$ increase, which is typical very large near quarter filling.
The spinon pairing temperature sets up as the superconducting temperature, $T_c=T_{\text{BCS}}$.
While there exist pseudogap phase in high-$T_c$ cuprates \cite{kotliar1988,proust2019highTc}, such phase is absent in the RNO materials since $T_{\text{BCS}}<T_{\text{BEC}}$.

{\bf RE element substitution}
Under the RE element substitution from La to Sm, the lattice constants decrease and the overlaps of neighbor electronic orbitals increase.
Both the intra-layer $3d_{x^2-y^2}$ orbital hopping and inter-layer $3d_{z^2}$ one will gradually increase.
On the contrary, the on-site Hubbard $U$ interaction does not changes much.
We adopt the hopping strength and Hubbard $U\simeq 4$eV from the DFT calculations in the Ref.~\cite{zhang2023trends} for different RE elements.
The effective inter-layer and intra-layer AFM spin super exchange are obtained from Eq.~(\ref{eq:Jcoupling}), as depicted in Fig.~(\ref{fig:REtJ}).
$J_{\perp}$ and $J_{\parallel}$ increase from La to Sm, while their ratio nearly unchanged, $J_{\perp}/J_{\parallel}\approx 1.65-1.70$.
Especially, for element substitution of Pm or Sm, the hopping strength and spin exchange get stronger much. 
Upon the RE element substitution, all the relevant energy scales become enhanced simultaneously, raising up the possibility of increased $T_c$.

\begin{figure}[t]
\includegraphics[width=1.0\linewidth]{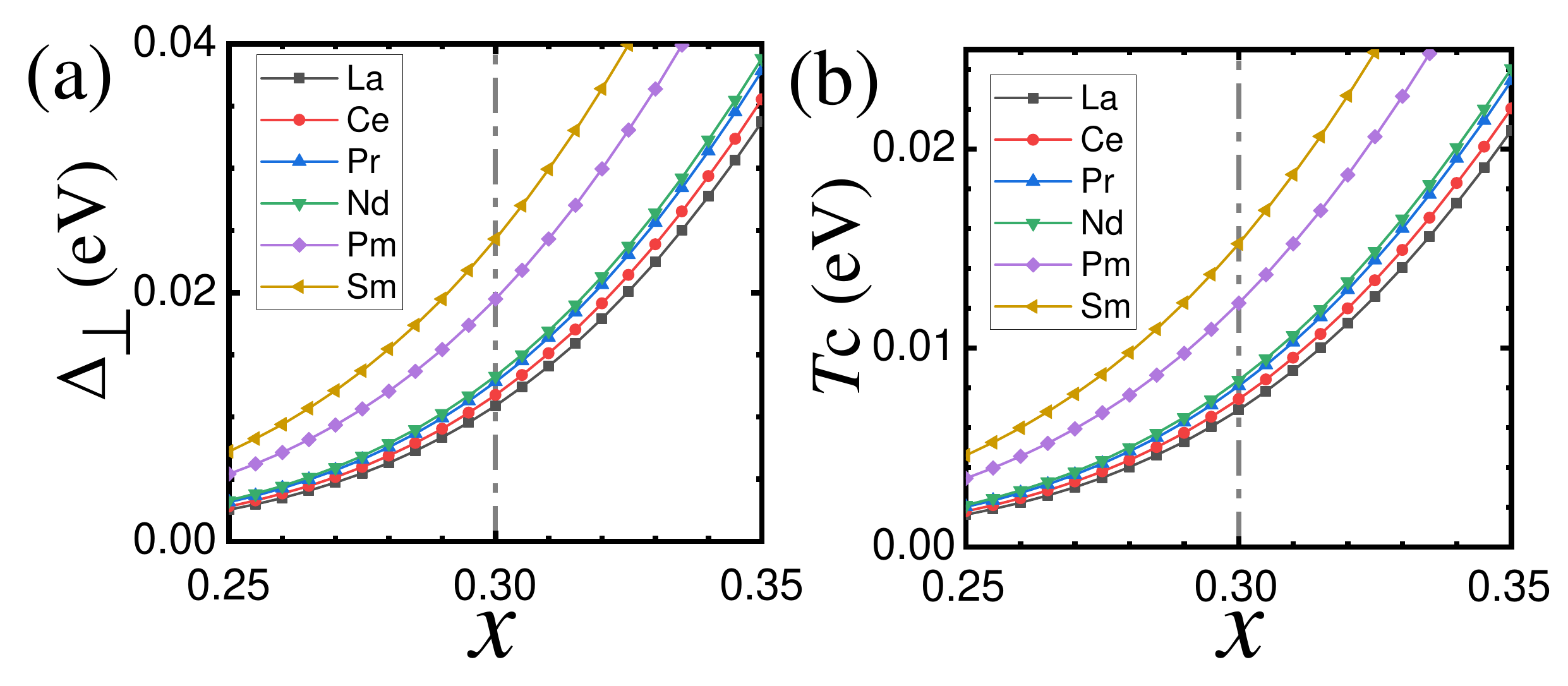}
\caption{(a). Inter-layer pairing gap $\Delta_{\perp}$ versus filling level $x$ under element substitution for R$_3$Ni$_2$O$_7$ (R=rare-earth element from La to Sm).
The place of physical filling $x\approx 0.3$ is shown in a dashed-dotted line.
The superconducting gap increases as the filling level grows.
(b). Superconducting transition temperature $T_c$ versus filling level $x$ under element substitution for R$_3$Ni$_2$O$_7$ (R=rare-earth element).
From La to Sm element substitution, the pairing strength $\Delta_{\perp}$ and the superconducting $T_c$ in RNO series increase simultaneously.
}
\label{fig:REDeltaTc}
\end{figure}

In the relevant parameter regime, 
an inter-layer $s$-wave pairing with high $T_c$ is deduced for RNO materials, similar as in LNO \cite{lu2023bilayertJ}.
The superconducting state is dominated by the inter-layer $3d_{x^2-y^2}$ orbital $s$-wave pairing $\Delta_{\perp}$, where the intra-layer one $\Delta_{\mu}^{(\alpha)}$ ($\mu=x,y$ and $\alpha=1,2$) nearly vanishes.
Such a strong inter-layer pairing is a direct consequence of the large inter-layer superexchange $J_{\perp}$ compared to intra-layer one $J_{\parallel}$.
The inter-layer pairings $\Delta_{\perp}$ from La to Sm elements grow up as the filling level increases, as depicted in Fig~\ref{fig:REDeltaTc}(a).
The increased inter-layer pairing indicates possible higher BCS pairing temperature from element substitution.

The strong inter-layer superexchange $J_{\perp}$ enhances the transition temperature dramatically, 
leading to the possible high $T_c$ superconductivity in the RNO materials.
The superconducting $T_c$ for RNO is numerically estimated as a function of the filling level $x$, as depicted in Fig.~\ref{fig:REDeltaTc}(b).
As $x$ increases from quarter filling ($x=0.25$), the transition temperature tends to increases monotonically.
At the same time, from La to Sm element substitution, the transition temperature is enhanced for the same filling level and get a significant jump from element substitution of La to Pm or Sm.
Such a strong enhancement in the temperature $T_c$ results from the enhancement of the energy scales for Pm and Sm element, as already be seen clearly in Fig.~(\ref{fig:REtJ}).
Moreover, the ratio between the pairing gap $\Delta_{\perp}$ and superconducting $T_c$ is approximately $\Delta_{\perp}/T_c\approx 0.4/0.25=1.6$,
which is quantitatively consistent with the prediction of BCS-type theory.

At the physical filling $x\approx 0.3$, 
the superconducting $T_c$ nearly doubles from La to Sm element.
It is expected that Sm$_3$Ni$_2$O$_7$ is the optimal material with the largest $T_c$ in the RNO material series.

{\bf Discussion}
The superconducting pairing and $T_c$ in the RNO materials are explored based on the bilayer single $3d_{x^2-y^2}$ orbital $t$-$J_{\parallel}$-$J_{\perp}$ model.
The numerical simulation shows that Sm$_3$Ni$_2$O$_7$ could exhibit the largest superconducting $T_c$.
The scenario proposed here differs from previous weak coupling analysis based on RPA\cite{zhang2023trends}.
There, the pairing strength as well as superconducting temperature decrease from La to Sm element substitution, and LNO is already the optimal high-$T_c$ material in the RNO series.
In our strong coupling picture, the superconducting pairing is mainly generated from the effective inter-layer spin superexchange $J_{\perp}$.
$J_{\perp}$ naturally increases as the inter-layer hopping strength increases under RE element substitution, and the superconductivity would be enhanced.

As we emphasized in the previous work\cite{lu2023bilayertJ} and further confirmed in this work, 
the high-$T_c$ superconductivity is driven by the strong interlayer spin exchange $J_{\perp}$ for the $3d_{x^2-y^2}$ orbitals in the RNO candidates.
The inter-layer $s$-wave superconducting pairing in the $3d_{x^2-y^2}$ orbital is favored in the relevant parameter regime, and the intra-layer hopping severs as the mobile engine.
Such a situation is distinct from the $d$-wave pairing in the single-layer curprate \cite{kotliar1988} and multilayer cuprates where the inter-layer exchange $J_{\perp}$ is naturally smaller than $J_{\parallel}$ \cite{ubbens1994,kuboki1995,maly1996,nazarenko1996bilayer3d,medhi2009}.
The wavefunctions of $3d_{x^2-y^2}$ orbital nearly lie within the plane and the interlayer overlap between $3d_{x^2-y^2}$ orbitals is small.
The strong $J_{\perp}> J_{\parallel}$ is hard to be directly generated from the much smaller inter-layer hopping of single $3d_{x^2-y^2}$ orbitals compared to the intra-layer one.
In the RNO materials, inter-layer overlap of $3d_{z^2}$ orbital wavefunctions persists strong inter-layer hopping.
Strong $J_{\perp}$ could be realized for $3d_{z^2}$ orbitals and is further transmitted to $3d_{x^2-y^2}$ orbitals under strong Hund's coupling.

Although we focus on the reduced single $3d_{x^2-y^2}$ orbital model, the effect of hidden $3d_{z^2}$ degrees of freedom plays an important role.
$3d_{z^2}$ electrons are nearly localized within the plane and has strong inter-layer hopping, inducing the strong inter-layer super-exchange $J_{\perp}$.
There also exist inter-layer pairings for the almost half-filled $3d_{z^2}$ orbitals.
However, the inter-layer $3d_{z^2}$ pairings are also nearly localized.
Even worse, $3d_{z^2}$ orbitals are nearly half-filled and holon condensation temperature is quite low due to its low density.
$3d_{z^2}$ orbitals do not contribute to the superconducting transport behavior directly.
On the contrary, the Hund's rule combines the two $E_g$ orbitals into the spin-triplet states, 
and induces effective strong inter-layer couplings $J_{\perp}$ between $3d_{x^2-y^2}$ orbitals.

While the simple single orbital model characterizes the most relevant physics of the superconducting LNO material \cite{lu2023bilayertJ}, 
further theoretical work and experimental verification are necessary to understand the interplay between $3d_{x^2-y^2}$ and $3d_{z^2}$ orbitals.
Moreover, the nature of the superconducting in $3d_{x^2-y^2}$ and $3d_{z^2}$ orbitals is different from each other.
A comprehensive analysis takes into account both degrees of freedom will be helpful.
Possible chemical doping approach may alter the superconducting behavior and further increase the $T_c$.
We will leave these problems in the further works.

{\bf Conclusion}
In this work, we focus on the effect of RE element substitution to the high $T_c$ superconductivity in the RNO materials under pressure based on the strong coupling bilayer single $3d_{x^2-y^2}$ orbital $t$-$J_{\parallel}$-$J_{\perp}$ model.
For the RE element from La to Sm, the effective inter-layer and intra-layer spin superexchange increase as the relevant hopping parameters grow up.
In the relevant filling level, the superconducting $T_c$ obtained is enhanced under the element substitution.
Strikingly, $T_c$ nearly doubles from substitution of La to Sm element, which might be experimental realized in the future.
This works suggest that RE element substitution in RNO materials serves as an important approach to enhance the superconducting $T_c$, 
appealing for further experimental verification.

\noindent
{\bf Acknowledgments}
F.Y. is supported by the National Natural Science Foundation of China under the Grants No. 12074031, No. 12234016, and No. 11674025.
C.W. is supported by the National Natural Science Foundation of China under the Grants No. 12234016 
and No. 12174317. 
This work has been supported by the New Cornerstone Science
Foundation.

\twocolumngrid
\bibliography{references}

\end{document}